\documentclass[prl,nofootinbib,twocolumn]{revtex4-2}
\usepackage{graphicx, epsfig}
\usepackage{color}
\usepackage{mathrsfs}
\usepackage{bm}
\usepackage{amsmath,amssymb}
\usepackage[caption=false]{subfig}
\usepackage{hyperref}
\usepackage[normalem]{ulem}
\usepackage{mathtools}

\def	\bqt	{\begin{quote}}
\def	\eqt	{\end{quote}}

\newcommand{\be}{\begin{equation}}
\newcommand{\ee}{\end{equation}}
\newcommand{\ba}{\begin{eqnarray}}
\newcommand{\ea}{\end{eqnarray}}
\newcommand{\nn}{\nonumber}

\begin{document}

\title{Quantum Mechanics of Gravitational Waves}
\author{Maulik Parikh$^{1,2}$}
\author{Frank Wilczek$^{1,3,4,5}$}
\author{George Zahariade$^{1,2}$}
\affiliation{$^{1}$Physics Department, Arizona State University, Tempe, AZ 85287, USA}
\affiliation{$^{2}$Beyond Center for Fundamental Concepts in Science, Arizona State University, Tempe, Arizona 85287, USA}
\affiliation{$^{3}$Department of Physics, Stockholm University, Stockholm SE-106 91, Sweden}
\affiliation{$^{4}$Center for Theoretical Physics, Massachusetts Institute of Technology, Cambridge, Massachusetts 02139, USA}
\affiliation{$^{5}$Wilczek Quantum Center, Department of Physics and Astronomy, Shanghai Jiao Tong University, Shanghai 200240, China}

\begin{abstract}
\noindent For the purpose of analyzing observed phenomena, it has been convenient, and thus far sufficient, to regard gravity as subject to the deterministic principles of classical physics, with the gravitational field obeying Newton's law or Einstein's equations. 
Here we treat the gravitational field as a quantum field and determine the implications of such treatment for experimental observables. We find that falling bodies in gravity are subject to random fluctuations (``noise") whose characteristics depend on the quantum state of the gravitational field. We derive a stochastic equation for the separation of two falling particles. Detection of this fundamental noise, which may be measurable at gravitational wave detectors, would vindicate the quantization of gravity, and reveal important properties of its sources.
\end{abstract}

\date{\today}
\maketitle

\section*{Introduction}

\noindent The behavior of objects falling freely under the influence of gravity is commonly described by Einstein's general theory of relativity, with the curvature field treated as a classical background. Individual test particles follow the geodesic equation while the separations of pairs of test particles obey the geodesic deviation equation. These are deterministic equations, befitting the classical theory from which they are derived. But the fundamental laws of physics are quantum-mechanical and, in the context of gravity, we expect the spacetime metric to be a quantum field.  To take this into account, a different framework is required.  

Here we present a formalism for calculating the effect on falling bodies due to the quantization of the gravitational field. We find that the dynamics of the separation of a pair of falling particles is no longer deterministic, but probabilistic, being acted on by a novel stochastic force. Specifically, we find that the separation of the two particles now obeys a Langevin-like stochastic equation containing a random fluctuation term, or noise~\cite{Parikh:2020nrd} (as is further discussed in the Supplemental information and more fully in~\cite{longpaper}). This provides the quantum generalization of the classical geodesic deviation equation. Our result applies also to a single object falling in the gravitational field of a heavier, fixed mass. Thus, an apple falling in Earth's gravity, say, would not fall straight down but would be subject to minute quantum jitters, which can be regarded heuristically as arising due to the bombardment of the apple by gravitons.

This effect is potentially measurable at gravitational wave detectors. We can model the mirrors of an arm of a gravitational wave interferometer~\cite{Abbott:2016blz,Audley:2017drz} as two freely-falling particles, and couple them to a quantized weak gravitational field. Then, using our formalism (which is based on the Feynman-Vernon influence functional), the effect on the separation of the mirrors can be calculated; the result is that the mirror separation is subject to quantum-gravitational noise. Moreover, the unusual power spectrum of this noise can allow it to be distinguished from many other sources of noise that gravitational wave interferometers are susceptible to~\cite{TheLIGOScientific:2016agk}. The statistical properties of the noise depend on the quantum state of the gravitational field, and we have calculated it explicitly for several classes of states. We estimate that the noise is unmeasurably small for coherent states, which are minimum-uncertainty quantum states that most closely resemble classical gravitational waves. However, there are theoretically predicted, though as yet unobserved phenomena, involving evaporation of black holes and exotic phases in the early universe, 
wherein quantum aspects of gravitational radiation play a central role. For the corresponding quantum states we find that the noise can be significantly enhanced. In particular, in squeezed states the noise can be enhanced exponentially in the squeezing parameter. Detection of this fundamental noise would provide experimental evidence for the quantization of gravity. Finally, we also discuss the connection between features of the radiation sources and the quantum nature of the radiation field.   

\section*{Analysis}

\noindent We are interested in how a pair of free-falling particles responds to a quantized gravitational field (compare~\cite{AmelinoCamelia:1998ax,AmelinoCamelia:1999gg,Dyson:2013jra,Verlinde:2019xfb,Verlinde:2019ade,Guerreiro:2019vbq}). We refer to the pair as a detector since the two mirrors at the ends of the arm of a gravitational wave interferometer can be idealized as two free-falling massive particles in a weak gravitational field. (This description would hold more literally for a space-based interferometer.) Suppose that the initial state of the gravitational field is $|\Psi \rangle$. As the field interacts with the detector, its quantum state changes because the detector generically both absorbs and emits gravitons through spontaneous and stimulated emission; the final field state $|f \rangle$ is a priori unknown. We would like to know the transition probability for the detector to go from state $|A \rangle$ to state $|B \rangle$ in time $T$. Since we do not measure the final state of the gravitational field, we must sum over $|f \rangle$. Thus we wish to calculate
\be
P_{\Psi}(A \to B)  =  \sum_{|f\rangle} |\langle f, B | \hat{U}(T) | \Psi, A \rangle |^2 \ , \label{prob}
\ee
where $|a, b \rangle \equiv |a \rangle \otimes |b \rangle$ and $\hat{U}$ is the unitary time-evolution operator for the combined gravitational field+detector system. 

\begin{figure}
\centering
\includegraphics[width=0.45\textwidth]{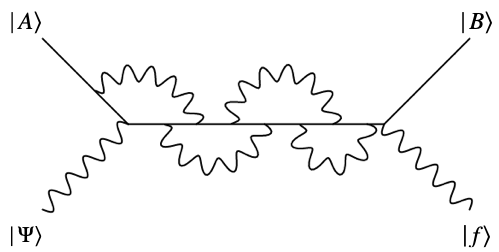}
\caption{A generic Feynman diagram representing an elementary process involved in the transition probability in equation~\eqref{prob}. Solid lines represent the detector while wiggly ones represent gravitons. Notice that, given equation~\eqref{action}, the only vertices allowed are graviton-detector-detector vertices which eliminates the possibility of pure graviton loops. Moreover, since the detector is ultimately expected to behave classically, we also disregard pure detector loops.}
\label{feynman}
\end{figure}

To go further, we need a more detailed description of the observed degrees of freedom; it is significant to focus on observables, because the natural variables include unphysical gauge dependence. Let the geodesic separation of the two particles be $\xi (t)$. The dynamics of the combined system of gravity and the two particles is described by the Einstein-Hilbert action minimally coupled to the actions of the two non-relativistic particles. The weakness of the gravitational field allows us to expand the Einstein-Hilbert action to second order in the metric perturbation $h_{\mu \nu}$. Then
\ba
S &=&-\frac{c^4}{64 \pi G}\int d^4x\ \partial_\mu {h}_{ij}\partial^\mu {h}^{ij}\nn\\
&&+ \int dt \frac{1}{2} m_0 \left (\delta_{ij} \dot{\xi}^i \dot{\xi}^j -\dot{{h}}_{ij} \dot{\xi}^i \xi^j \right ) \ . \label{action}
\ea
In this expression, we are left with only a single degree of freedom for the two particles: their separation. Thus, our results apply also to the case of a single particle subject to the gravity of a heavier, fixed mass. We can now evaluate the amplitudes in equation~\eqref{prob} in the path integral formulation derived from the action in equation~\eqref{action}, where $|A \rangle$ and $|B \rangle$ are the initial and final states of $\xi$; see Fig.~\ref{feynman}. Thus we have
\be
P_{\Psi}(A \to B) = 
{\cal I}_{A,B}\hspace{-1mm} \int {\cal D}\xi \, {\cal D}\xi' e^{\frac{i}{\hbar} \int dt \frac{m_0}{2}  (\dot{\xi}^2 - \dot{\xi}'^{2})} F_{\Psi}[\xi,\xi'] \ .
\label{transfinal}
\ee
This expression can be understood as follows. The double path integral reflects the fact that this is a probability, rather than an amplitude. The factor ${\cal I}_{A,B}$ contains integrals over the initial and final wave functions of $\xi$, and will play no further role. In the exponent, we recognize the non-relativistic action for a free particle. Equation~\eqref{transfinal} gives us an effective theory for the particle separation $\xi$, in which the effects of its coupling to the quantized gravitational field have been taken into account. Crucially, all the effects of the quantized gravitational field are formally captured by the functional $F_{\Psi}[\xi,\xi']$, known as the Feynman-Vernon influence functional \cite{Feynman:1963fq,Calzetta:1993qe,Hu:1999mm}.

To evaluate the influence functional, we write the gravitational field state $|\Psi \rangle$ as a tensor product of single-mode states: $|\Psi \rangle = \bigotimes_{\vec{k}}|\psi_{\vec{k}}\rangle$. For a weak gravitational field, linearity allows us to treat the problem mode-by-mode and then sum over modes. Then $F_{\Psi}[\xi,\xi'] = \prod_{\vec{k}} F_{\psi_{\vec{k}}}[\xi,\xi']$, where $F_{\psi_{\vec{k}}}[\xi,\xi']$ is the influence functional for a single mode of the gravitational field. To compute this, we decompose the metric perturbation in Fourier modes. Let the mode of wave number $\vec{k}$ have angular frequency $\omega_{\vec{k}}$ and amplitude $q_{\vec{k}}(t)$. If we now, for simplicity, assume that the perturbation propagates orthogonally to the line joining the particles, and if we further restrict to a single polarization, then we find that the gravitational part of the action in equation~\eqref{action}, for a single mode of the gravitational field is
\be
S^{\vec{k}}_\xi = \int dt \left ( \frac{1}{2} m (\dot{q}_{\vec{k}}^{2} - \omega_{\vec{k}}^2  q_{\vec{k}}^2) - g \dot{q}_{\vec{k}} \dot{\xi} \xi  \right ) \, . \label{Somega}
\ee
Here $g = \frac{m_0}{2 \sqrt{\hbar G/c^3}}$ is a coupling constant; $m$ is a mass introduced for dimensional reasons and will drop out of all physical expressions after integration over modes.

The model can be made more realistic by including the other polarization, as well as by properly accounting for various trigonometric factors that would arise for modes that are incident from different directions; here we neglect those complicating refinements to focus on the core problem.

Notice that the single-mode action in equation~\eqref{Somega} describes a simple harmonic oscillator of angular frequency $\omega_{\vec{k}}$ coupled to an external source $\xi$, via a cubic, derivative interaction. We can readily quantize this simple action to evaluate the influence functional; the calculation can be performed exactly, without invoking perturbation theory, because the Lagrangian is quadratic in $q$. The associated quantum Hamiltonian reads $\hat{H}^{\vec{k}}_\xi = \frac{(\hat{p}_{\vec{k}} + g \xi \dot{\xi})^2}{2m} + \frac{1}{2} m \omega_{\vec{k}}^2 \hat{q}_{\vec{k}}^2$. Correspondingly, the expression for the single-mode influence functional is $F_{\psi_{\vec{k}}}[\xi, \xi'] = \langle \psi_{\vec{k}} | \hat{U}^{\vec{k} \dagger} _{\xi'} (T) \hat{U}^{\vec{k}}_\xi (T) | \psi_{\vec{k}} \rangle$, where $\hat{U}^{\vec{k}}_\xi$ is the unitary time-evolution operator obtained from $\hat{H}^{\vec{k}}_\xi$. Repeated application of the Baker-Campbell-Hausdorff formula then yields
\be
F_{\psi_{\vec{k}}}[\xi,\xi']
= F_{0_{\vec{k}}}[\xi,\xi'] \langle \psi_{\vec{k}} | e^{-W^* \hat{a}^\dagger} e^{W \hat{a}} | \psi_{\vec{k}} \rangle \ ,
\label{geninffunc}
\ee
where $W$ and $F_{0_{\vec{k}}}[\xi,\xi']$ are readily calculated functions of $\xi$ and $\xi'$.
Equation~\eqref{geninffunc} is useful computationally because the exponents are expressed in terms of ladder operators, $\hat{a}$ and $\hat{a}^\dagger$, whose operation on $|\psi_{\vec{k}} \rangle$ can be calculated for many classes of states.  In particular, equation~\eqref{geninffunc} can be evaluated in closed form for vacuum, coherent, thermal, and squeezed states.

Having obtained the influence functional for a single mode of the gravitational field in an initial state $|\psi_{\vec{k}} \rangle$, we are now ready to sum over modes to find the total influence functional of the gravitational field from the product of single-mode influence functionals. The result will depend on the quantum state of the gravitational field. As a basic example, suppose the gravitational field is in its vacuum state. Performing the mode sum, we find in particular that
\ba
\ln |F_{\rm vac} | &=& -\frac{m^2_0}{32\hbar^2}\int_0^T\hspace{-2mm}\int_0^T\hspace{-2mm} dt\,dt'\,A(t-t')\times\nn\\ &&\hspace{5mm}\left(X(t)-X'(t)\right)\left(X(t')-X'(t')\right) \label{lnF} \ ,
\ea
where $X = \frac{d^2}{dt^2} (\xi^2)$, $X' = \frac{d^2}{dt^2} (\xi'^2)$, and $A(t-t')$ is a known integral. Feynman and Vernon realized that whenever $\ln |F|$ is quadratic in $X-X'$, then $|F|$ can be rewritten in a very suggestive manner, as a statistical average over an auxiliary function $N(t)$: 
\be
|F_{\rm vac} | = \left < 
\exp \left ( \frac{i}{\hbar}\int_0^T dt\frac{m_0}{4}N(t)\left(X(t)-X'(t)\right) \right ) \right >_N \ . \label{Fvac}
\ee
Here the function $N(t)$ obeys a Gaussian probability distribution with a stationary auto-correlation function $A(t-t')$. Thus the effect of coupling to the quantum gravitational vacuum, which is fully encoded in $F_{\rm vac}$, is to introduce stationary stochastic noise in the detector. As we will see, this creates fluctuations in the length of the arm. All the statistical properties of the fluctuations, such as the standard deviation, can be obtained from the auto-correlation function, $A$. In contrast, we find that the phase of $F_{\rm vac}$ has a different structure, which is responsible for dissipative effects. For a coherent state, corresponding to a classical gravitational wave $h(t)$, the phase of the influence functional also contains a term $\frac{i}{\hbar}\int_0^T dt\frac{1}{4}m_0 h(t)\left(X(t)-X'(t)\right)$.

\section*{Quantum geodesic deviation}

\noindent We have so far focused on the Feynman-Vernon influence functional for a detector coupled to a gravitational field, treating both the detector and the field quantum-mechanically. Since realistic detectors are well-approximated as classical, it is appropriate to exploit that simplification. In our expression for the transition probability, equation~\eqref{transfinal}, we identify the classical paths, which dominate the integral, as those which render the phases stationary. This leads to an effective {\it stochastic\/} equation of motion for the separation of the masses or, equivalently, for the arm length of a gravitational wave detector. In the presence of a classical gravitational wave $h(t)$, we find
\be
\ddot{\xi}(t)-\frac{1}{2}\left[\ddot{h}(t)+\ddot{N}(t)-\frac{m_0G}{c^5}\frac{d^5}{dt^5}\xi^2(t)\right]\xi(t)=0\ .
\label{langevineq}
\ee
Thus the relative acceleration of the two masses, $\ddot{\xi}$, depends on three terms. Each term in this equation, which extends the  geodesic deviation equation of general relativity to the case where the spacetime metric is treated as a quantum field, has intuitive meaning.  The first represents the usual tidal acceleration due to the passing of a classical gravitational wave, $h$; this is the effect that has been famously measured at LIGO. The last term is the gravitational counterpart of the dissipative Abraham-Lorentz term in electromagnetism; it is the gravitational radiation reaction~\cite{Thorne:1969rba,Burke:1970wx,Chandra,Mino:1996nk}. Although it is likely to be of little experimental consequence, it is nevertheless notable that such a term arises from a well-behaved quantum theory.  (As will be reported elsewhere, this approach to radiation reaction avoids the notorious pathologies that arise from too literal interpretation of the Abraham-Lorentz equation~\cite{Feynman:1963uxa}). Most importantly, our equation contains a noise term, $\ddot{N}$. Thus our equation, while classical, is stochastic rather than deterministic~\cite{Calzetta:1993qe,Hu:1999mm}. It is reminiscent, mathematically, of the Langevin equation used to describe Brownian motion. 

\begin{figure}
\centering
\includegraphics[width=0.45\textwidth]{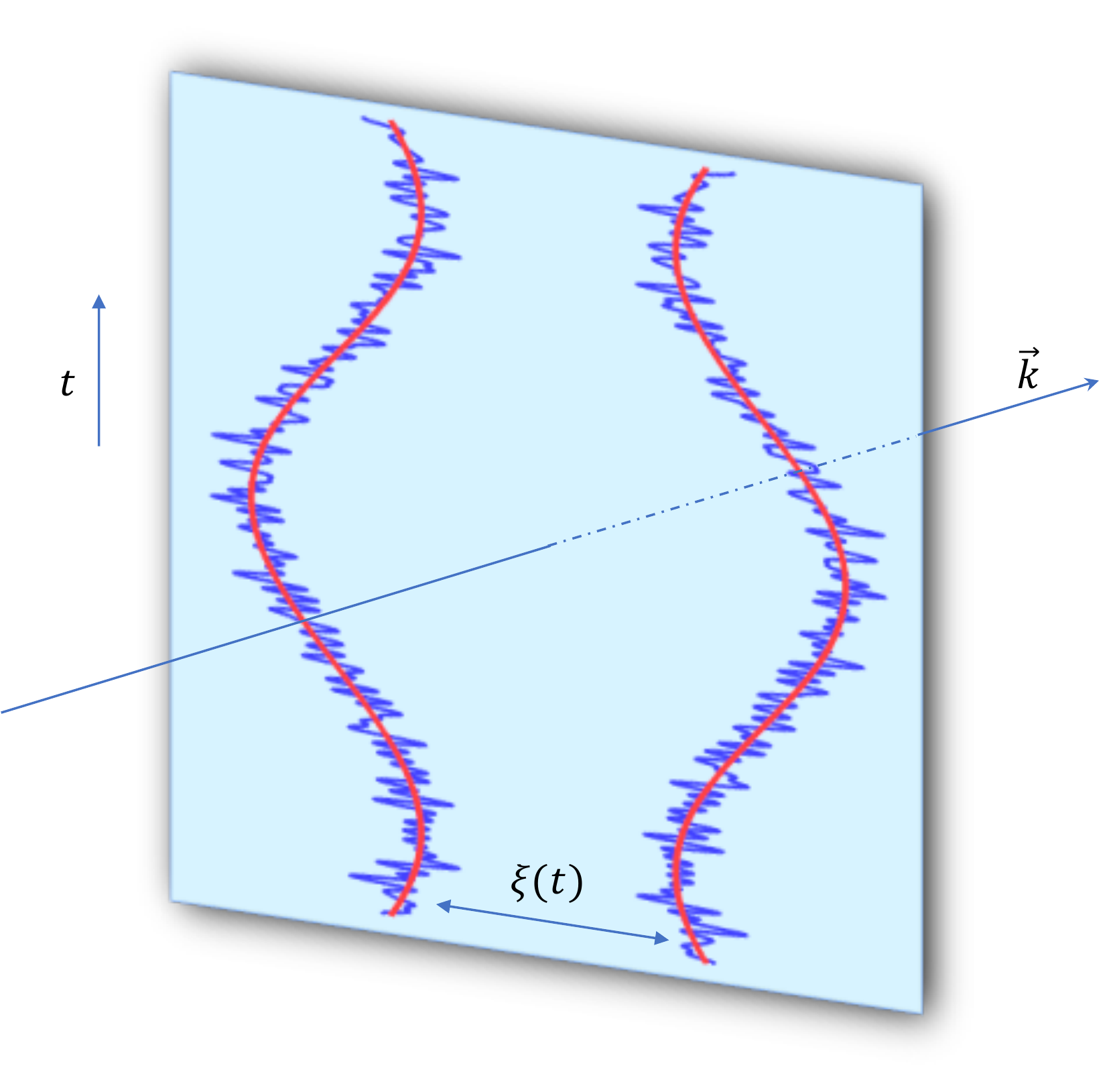}
\caption{Schematic spacetime diagram of the motion of the endpoints of the detector in the presence of one polarization of a quantized gravitational wave propagating along $\vec{k}$. On average, each particle traces an inertial trajectory, following a geodesic of the classical spacetime. The physical separation $\xi(t)$ of the two particles obeys, on average, the geodesic deviation equation. The effect of quantization of the gravitational field is to introduce a random fluctuation on top of this classical motion.}
\label{fluctuations}
\end{figure}

We have calculated the statistical properties of this noise, specifically the power spectrum, $S(\omega)$, in several cases. For the vacuum and for coherent states we find that, for low frequencies, the power spectrum behaves like $S = 4 G \hbar \omega/c^5$. For thermal states at temperature $T$, we find $S = \frac{4 G \hbar \omega}{c^5} \coth \left (\frac{\hbar \omega}{2 k_B T} \right )$. Strikingly, if the gravitational field is in a squeezed state with squeezing parameter $r$, we find that the noise also contains a non-stationary piece; focusing on the stationary part, we find $S = \sqrt{\cosh 2r} \, 4 G \hbar \omega/c^5$, which means that the quantum-induced fluctuations of the detector arm length can be exponentially enhanced.

\section*{Discussion}

\noindent We have shown that a pair of freely-falling particles in a quantized gravitational field obeys the stochastic equation~\eqref{langevineq}, rather than the classical geodesic deviation equation; see Fig.~\ref{fluctuations}. Let us estimate the size of the fluctuations, neglecting the radiation reaction term. Then $\xi(t) \approx (1+ \frac{1}{2} (h + N)) \xi_0$, where $\xi_0$ is the equilibrium length of the arm; for LIGO, $\xi_0 \approx 4 \, {\rm km}$, while for LISA, $\xi_0 \approx 10^6 \, {\rm km}$. As the formula makes clear, the fluctuations of $N$ induce fluctuations in $\xi$. We find that $\langle \xi \rangle = (1+ \frac{h(t)}{2}) \xi_0$, with a variance $\sigma^2 = \frac{\xi_0^2}{4} \langle N^2 \rangle = \frac{\xi_0^2}{4} A(0)$. Although $A(0) = \frac{1}{\pi}\int_0^\infty d \omega S(\omega)$ is formally divergent, the size of fluctuations is nevertheless finite because limits on detector sensitivity impose a cut-off, $\omega_{\rm max}$, on the frequency integral; for LIGO, $\omega_{\rm max} \sim 1 \, {\rm kHz}$ while for LISA, $\omega_{\rm max} \sim 0.1 \, {\rm Hz}$. With these numerical values, the amplitude of the fluctuations in the vacuum state, $\sigma \sim l_P \xi_0 \, \omega_{\rm max}/c$, is roughly a Planck length, $l_P$, and therefore completely unobservable; essentially, the same amplitude is obtained also when the gravitational field is in a coherent state corresponding to a classical gravitational wave propagating in the vacuum~\cite{Dyson:2013jra}.

However, the amplitude of the fluctuations can be enhanced for non-coherent states of the gravitational field. As representative examples, we may consider states formed by the action of displacement operators representing the classical field $h$ acting not upon the vacuum, but upon thermal or squeezed states. For thermal states at temperature $T$, we find an enhancement by a factor $\sqrt{\frac{2 k_B T}{\hbar \omega_{\rm max}}}$. Such thermal states could arise, for example, through the isotropic cosmic gravitational wave background ($T \sim 1 K$) or as a result of the Hawking evaporation of black holes~\cite{Hawking:1974rv}. In the latter case, the temperature can be very high, but there is an additional suppression due to the localization of the source. Squeezed vacua can arise in inflationary cosmology~\cite{Grishchuk:1990bj,Albrecht:1992kf,Koks:1996ga}. As noted above, we expect the amplitude of fluctuations to be enhanced exponentially in the squeezing parameter $\sigma \sim e^{\frac{r}{2}} l_P \xi_0 \, \omega_{\rm max}/c$.

It is enlightening to compare the quantum nature of electromagnetic and gravitational radiation fields, for known and contemplated sources.   It is almost always appropriate to treat the coupling of the electromagnetic fields to its sources as linear.   When one has linear coupling of a radiation field to dynamical degrees of freedom which are described, to a good approximation, as deterministic and only weakly perturbed by the radiation, then the radiation field will be well described by a coherent quantum state~\cite{Glauber:1963tx}. This is the case for most radio and microwave sources, and for lasers.   When the sources themselves are stochastic, one obtains a stochastic mixture of coherent states.   This is the case for the most common (quasi-thermal) higher frequency sources.  With special techniques, e.g. using nonlinear crystals, one can construct sources whose coupling to the electromagnetic field is quadratic, leading to squeezed states.    The default ``classical'' treatment of gravitational radiation, which corresponds to coherent states, is appropriate when the sources are governed by approximately deterministic dynamics involving weak linear coupling to the gravitational field. This is often an appropriate default, e.g. in describing slow orbital decay of large bodies.  It does not apply to Hawking radiation, which is a quasi-thermal quantum process, nor to its cosmological analogues nor, more speculatively, to phase transitions in the early universe. During the late stages of black hole mergers, the approximation of treating gravitational radiation as a weak linear perturbation is not appropriate either, despite the deterministic nature of the dynamics.  Here one can expect to encounter effects we might call {\em molding\/} of the quantum radiation state, which go beyond (quadratic) squeezing. 

\section*{Acknowledgments}

\noindent We thank Paul Davies, Bei-Lok Hu, Phil Mauskopf, Siddharth Morampudi, Igor Pikovski, and Tanmay Vachaspati for conversations. During the course of this work, MP and GZ were supported in part by John Templeton Foundation grant 60253. GZ also acknowledges support from the Foundational Questions Institute and Moogsoft. FW is supported in part by the U.S. Department of Energy under grant DE-SC0012567, by the European Research Council under grant 742104, and by the Swedish Research Council under contract 335-2014-7424.

\end{document}


\title{Supplemental information to ``Quantum Mechanics of Gravitational Waves''}
\author{Maulik Parikh$^{1,2}$}
\author{Frank Wilczek$^{1,3,4,5}$}
\author{George Zahariade$^{1,2}$}
\affiliation{$^{1}$Physics Department, Arizona State University, Tempe, Arizona 85287, USA}
\affiliation{$^{2}$Beyond Center for Fundamental Concepts in Science, Arizona State University, Tempe, Arizona 85287, USA}
\affiliation{$^{3}$Department of Physics, Stockholm University, Stockholm SE-106 91, Sweden}
\affiliation{$^{4}$Center for Theoretical Physics, Massachusetts Institute of Technology, Cambridge, Massachusetts 02139, USA}
\affiliation{$^{5}$Wilczek Quantum Center, Department of Physics and Astronomy, Shanghai Jiao Tong University, Shanghai 200240, China}

\date{\today}
\maketitle

\subsection*{Derivation of Equation~\eqref{action}}

We put one of the two freely-falling particles on-shell by, for example, making it very heavy (alternatively, we could discard the center of mass motion of the system). Next, we introduce Fermi normal coordinates associated with the geodesic of the heavy particle such that the coordinates of the other particle are $X^\mu = (t, \xi^i)$. In Fermi normal coordinates, the metric reads
\ba
g_{00}(t,\xi)&=&-1-R_{i0j0} (t,0) \xi^i\xi^j + O(\xi^3) \nn \ ,\\
g_{0i}(t,\xi)&=&-\frac{2}{3}R_{0jik} (t,0) \xi^j\xi^k+O(\xi^3) \nn \ ,\\
g_{ij}(t,\xi)&=&\delta_{ij}-\frac{1}{3}R_{ikjl} (t,0) \xi^k\xi^l+O(\xi^3)\nn\ .
\ea
Inserting this into the relativistic action $-m_0 \int d \tau \sqrt{-g_{\mu \nu} \dot{X}^\mu \dot{X}^\nu}$, using reparameterization invariance to set $\tau = t$, and expanding the square root to obtain the non-relativistic limit we obtain $-m_0 \int dt (1 + \frac{1}{2} R_{i0j0} (t,0) \xi^i \xi^j - \frac{1}{2} \delta_{ij} \dot{\xi}^i \dot{\xi}^j)$. We now exploit the fact that this expression is unchanged when switching from Fermi normal coordinates to transverse-traceless gauge, to first order in the metric perturbation $h$. Since $R_{i0j0}(t,0) = - \frac{1}{2} \ddot{h}_{ij}(t,0)$ in transverse-traceless gauge, we obtain the second integral in equation~\eqref{action}. Finally, we can express the quadratic part of the Einstein-Hilbert action as the first integral in equation~\eqref{action}.

\subsection*{Derivation of Equation~\eqref{transfinal}}
Starting from equation~\eqref{prob}, we have
\ba
P_{\Psi}(A \to B) & = & \sum_{|f\rangle} \langle \Psi, A  | \hat{U}^\dagger(T) | f , B \rangle \langle f, B | \hat{U}(T) | \Psi, A \rangle \nn \\
& = &  \int d \xi_i d \xi'_i d \xi_f d \xi'_f \phi^*_A(\xi'_i)  \phi_B(\xi'_f) \phi^*_B(\xi_f)  \phi_A(\xi_i) \ \sum_{|f\rangle} \langle \Psi,  \xi'_i | \hat{U}^\dagger(T) | f , \xi'_f \rangle \langle f, \xi_f | \hat{U}(T) | \Psi , \xi_i \rangle \nn\ ,
\ea
where $\phi_A(\xi) = \langle \xi | A \rangle$, etc. are the position-space wavefunctions. We can write the amplitudes heuristically as Feynman path integrals:
\be
\langle f, \xi_f | \hat{U}(T) | \Psi , \xi_i \rangle = \underset{\substack{\xi(0) = \xi_i \\ \xi(T) = \xi_f}} \int {\cal D} \xi \int {\cal D} h \, e^{\frac{i}{\hbar} S} \nn\ ,
\ee
where the action, $S$, is given in equation~\eqref{action}. The action contains a term independent of $h$: $S = S_\xi + S_{h,\xi}$, where $S_\xi = \int dt \frac{1}{2} m_0 \dot{\xi}^2$. This can be pulled out of the path integral over $h$.
Defining $I_{A,B} = \int d \xi_i d \xi'_i d \xi_f d \xi'_f \phi^*_A(\xi'_i)  \phi_B(\xi'_f) \phi^*_B(\xi_f)  \phi_A(\xi_i)$, we arrive at equation~\eqref{transfinal}, where
\be
F_\Psi[\xi,\xi'] = \sum_{|f\rangle} \int {\cal D} h \, {\cal D} h' \, 
e^{\frac{i}{\hbar} (S_{h,\xi} - S_{h',\xi'})} \nn\ .
\ee
The boundary conditions on the path integrals correspond to an initial gravitational field state $|\Psi \rangle$ and a final field state $|f\rangle$.

\subsection*{Derivation of Equation~\eqref{Somega}}
We can perform a Fourier mode decomposition on $h$ in a finite box of side $L$:
\be
h_{ij}(t,\vec{x}) = \frac{1}{\sqrt{\hbar G/c^3}} \sum_{\vec{k},s} \qk(t) e^{i \vec{k} \cdot \vec{x}} \epsilon^s_{ij} (\vec{k}) \nn\ ,
\ee
where $\epsilon^s_{ij}$ is the polarization tensor, and $s = +, \times$.
We use orthonormality of the modes, the reality of $h$, and properties of the polarization tensor in transverse-traceless gauge to write $S_{h, \xi}$ as a sum over modes:
\be
S_{h, \xi} =  \int dt \sum_{\vec{k},s} \frac{1}{2} m \left (  \dot{q}_{\vec{k},s}^{ 2} - \omega_{\vec{k}}^2  \qk^2 \right ) - \int dt \frac{1}{2} m_0 \frac{1}{\sqrt{\hbar G/c^3}} \sum_{\vec{k},s} \dot{q}_{\vec{k},s} \epsilon^s_{ij}(\vec{k}) \dot{\xi}^i \xi^j \nn\ .  
\ee
Here $m \equiv \frac{L^3}{16 \pi \hbar G^2}$ is a constant introduced solely for dimensional reasons. Focusing for simplicity only on wave vectors along a single direction orthogonal to the separation of the particles, and restricting to a single polarization mode, we obtain $S_{h, \xi} = \sum_{\vec{k}} S^{\vec{k}}_\xi$, where $S^{\vec{k}}_\xi$ is given by equation~\eqref{Somega}.

\subsection*{Derivation of Equation~\eqref{geninffunc}} 
We can express the path integrals in $F$ in canonical form:
\be
\int Dh \, e^{\frac{i}{\hbar} S_{h,\xi}} = \langle f | \hat{U}_{\xi} (T) |\Psi \rangle \nn\ ,
\ee
where $\hat{U}_{\xi} = \bigotimes_{\vec{k}} \hat{U}^{\vec{k}}_{\xi}$ and $\hat{U}^{\vec{k}}_{\xi}$ is the single-mode evolution operator associated with the quantum Hamiltonian $\hat{H}^{\vec{k}}_\xi$. Then we find
\be
F_\Psi[\xi,\xi'] = \sum_{|f\rangle} \langle \Psi | \hat{U}^\dagger_{\xi'} (T) |f \rangle \langle f | \hat{U}_{\xi} (T) |\Psi \rangle = \langle \Psi | \hat{U}^\dagger_{\xi'} (T) \hat{U}_{\xi} (T) |\Psi \rangle = \prod_{\vec{k}} \langle \psi_{\vec{k}} | \hat{U}^{\vec{k} \, \dagger}_{\xi'} (T) \hat{U}^{\vec{k}}_{\xi} (T) |\psi_{\vec{k}} \rangle \nn\ .
\ee
On the right, we have expressed the field-theoretic influence functional as a product of single-mode influence functionals. We can evaluate those in ordinary quantum mechanics using standard techniques, as follows. We work in interaction picture, and express all operators in terms of ladder operators. We then repeatedly invoke the relation
$e^{\hat{A}} e^{\hat{B}} = e^{\hat{A}+\hat{B}} e^{\frac{1}{2} [\hat{A},\hat{B}]}$, a variant of the Baker-Campbell-Hausdorff formula that holds when $[\hat{A},\hat{B}]$ is a $c$-number. A lengthy but straightforward calculation then yields equation~\eqref{geninffunc}, with $W = \frac{ig}{\sqrt{8m\hbar\omega_{\vec{k}}}}\int_0^T dt\left(X(t)-X'(t)\right)e^{-i\omega_{\vec{k}} t}$, where $X = \frac{d^2}{dt^2} ({\xi}^2)$, etc., and
\be
F_{0_{\vec{k}}}[\xi,\xi']\equiv \exp\left[-\frac{g^2}{8m\hbar\omega_{\vec{k}}}\int_0^T\int_0^t dt\,dt'\left(X(t)-X'(t)\right)\left(X(t')e^{-i\omega_{\vec{k}}(t-t')}-X'(t')e^{i\omega_{\vec{k}}(t-t')}\right)\right] \nn\ . \label{vacinffunc}
\ee

\subsection*{Derivation of Equation~\eqref{lnF}} We can now sum over modes, using $\frac{1}{m} \sum_{\vec{k}} \to \frac{8 \hbar G^2}{\pi} \int \omega^2 d \omega$. Then, expanding the integrand in real and imaginary parts, we immediately arrive at equation~\eqref{lnF}, where the auto-correlation function is
\be
A(t-t') = \frac{4\hbar G}{\pi}\int_0^\infty d\omega\,\omega\cos(\omega(t-t')) \nn\ .
\ee
This integral is formally divergent but can be regulated.

\subsection*{Derivation of Equation~\eqref{Fvac}}
The absolute value of the influence functional can be expressed in a suggestive form as a statistical average by using an identity first proposed by Feynman and Vernon~\cite{Feynman:1963fq}:
\ba
&&\exp\left[-\frac{m^2_0}{32\hbar^2}\int_0^T\int_0^T dt\,dt'\,A(t,t')\left(X(t)-X'(t)\right)\left(X(t')-X'(t')\right)\right]\nn\\
&&\hspace{-6mm}
=\int {\cal D}N\exp\left[-\frac{1}{2}\int_0^T\int_0^Tdt\,dt'\,A^{-1}(t,t')N(t)N(t')+\frac{i}{\hbar}\int_0^T dt\frac{m_0}{4}N(t)\left(X(t)-X'(t)\right)\right]\nn\ .
\label{funcgauss}
\ea
The right-hand side can be interpreted as a statistical average over a random function $N(t)$, with a zero-mean Gaussian probability distribution, and defines equation~\eqref{Fvac}.

\subsection*{Derivation of Equation~\eqref{langevineq}} For a coherent state (corresponding to a classical gravitational wave), we can easily evaluate $F$. Putting together our previous expressions, we find
 \ba
P_{h}(A \to B)
&=& I_{A,B} \int {\cal D}\xi {\cal D}\xi' {\cal D}N\exp\left[-\frac{1}{2}\int_0^T\int_0^Tdt\,dt'\,A^{-1}(t-t')N(t)N(t')\right]\times\nn\\
&&\hspace{8mm}\exp\Biggr[\frac{i}{\hbar} \int_{0}^{T} dt \left\{\frac{1}{2} m_0 \left(\dot{\xi}^2 - \dot{\xi}'^{2}\right)+\frac{1}{4}m_0\left(h(t)+N(t)\right)\left(X(t)-X'(t)\right)\right\}\nn\\
&&\hspace{12mm}-\frac{im_0^2 G}{8\hbar}\int_0^Tdt\,\left(X(t)-X'(t)\right)\left(\dot{X}(t)+\dot{X}'(t)\right)
\Biggr]\nn\ ,
\ea
where in the last line we have also included the phase of $F$. The dependence on $h(t)$ arises by applying equation~\eqref{geninffunc} on the coherent state. In the stationary phase approximation, the paths that dominate the probability obey the equation $\ddot{\xi}-\frac{1}{2}\left[\ddot{h}+\ddot{N}-\frac{m_0G}{2}\left(\dddot{X}+\dddot{X}'-\ddot{X}+\ddot{X}'\right)\right]\xi=0$, as well as a similar equation with $\xi$ and $\xi'$ interchanged. Setting $\xi = \xi'$, we finally obtain equation~\eqref{langevineq}.

\makeatletter\@input{xx.tex}\makeatother